\documentclass[letterpaper, 10 pt, conference]{ieeeconf}  

\IEEEoverridecommandlockouts                              

\usepackage{graphics}       
\usepackage{epsfig}         
\usepackage{times}          
\usepackage{amsmath}        
\usepackage{amssymb}        
\usepackage{color}
\usepackage{array}
\usepackage{accents}

\usepackage{stackengine}
\usepackage{url}

\title{\LARGE \bf Inversion-free feedforward hysteresis control using Preisach model}

\author{Michael Ruderman
\thanks{M. Ruderman is with University of Agder, Department of Engineering Sciences, Norway.
Email: {\tt\small michael.ruderman@uia.no}}%
\thanks{\textcolor[rgb]{0.00,0.00,1.00}{Author's accepted
manuscript, IEEE ECC2023}} }

\begin{document}

\maketitle
\thispagestyle{empty}
\pagestyle{empty}

\begin{abstract}
We introduce a new inversion-free feedforward hysteresis control
using the Preisach model. The feedforward scheme has a high-gain
integral loop structure with Preisach hysteresis operator in
negative feedback. This allows obtaining a dynamic quantity which
corresponds to the inverse hysteresis output, as the loop error
tends towards zero for a sufficiently high feedback gain. By
analyzing the loop sensitivity function with hysteresis that acts
as a state-varying phase lag, we demonstrate the achievable
bandwidth and accuracy of the proposed control method. Remarkable
fact is that the control bandwidth is theoretically infinite,
provided the Preisach operator in feedback can be implemented in a
way to ensure the $\mathcal{C}^0$ continuous hysteresis output.
Numerical control examples with the Preisach hysteresis model in
differential form are presented.
\end{abstract}

\bstctlcite{references:BSTcontrol}

\section{Introduction}
\label{sec:1}

Hysteresis phenomena occur in quite different physical and
technical systems, see e.g. \cite{BertMayer06}, and often require
an accurate compensation through dedicated control measures. From
a system and control viewpoint, the hysteresis can be seen as a
multi-valued quasi-static nonlinearity, affected by a nonlocal
memory. The latter implies that some part of the history of
previous states is retained and influences the current state of
the system. Often, the hysteresis appears inside of more complex
dynamic systems, so that the hysteresis output is not directly
measurable for a feedback control. It appears, for example, with
the magnetic flux density in controlled electromagnets or net
electrical charge in piezoelectric actuators. For systems without
hysteresis nonlinearity sensing, a pure feedforward control,
correspondingly compensation, often seems preferable over some
feedback control strategies.

In the control literature, a feedforward hysteresis control is
mostly associated with constructing a model-based (sometimes also
model-free) inverse (or its approximate) of the hysteresis. Some
earlier works on controlling the unknown hysteresis go back e.g.
to \cite{tao1995}. A more recent control framework with hysteresis
compensation was also reported in \cite{esbrook2012}. Apart from
that, a passivity-based stability and control of hysteresis were
addressed, still in a feedback manner, in \cite{gorbet2001}.
Another alternative compensation approach, which is worth to be
mentioned here since addressing the relevant phase shift
properties of a hysteresis system, was provided in
\cite{cruz2001}. A mixed recursive algorithm to control the output
remnant of a hysteresis system was presented in
\cite{vasquez2020}. An inverse multiplicative structure, that is
relying on internal model-based principle, was shown in
\cite{Janaideh2018} with use of the Prandtl-Ishlinskii (see e.g.
in \cite{Visit94}) hysteresis model in feedforwarding. However, an
artificial time-delay was necessary for bypassing an algebraic
loop of the proposed scheme. Multiple works on hysteresis
compensation with feedforward scheme have also been published in
the context of inverse mapping, e.g. \cite{krejci2001}, or more
generally inverse modeling of hysteresis systems to be controlled.
Due to a variety of (often ad-hoc and approximative) approaches,
and due to our focus on the Preisach \cite{Preis35} hysteresis
model, we will only refer to some of them. The theoretical
conditions for existence and properties of the inverse Preisach
operator were known since \cite{BrokateVisintin89}. However, to
the best of the author's knowledge, no closed analytical form of
the inverse Preisach operator suitable for feedforward control has
been reported so far. A remarkably fast computation of the
Preisach inverse, targeting the real-time implementation, was
reported e.g. in \cite{davino2005} and in the multiple following
works by the co-authors. The approach relies on the stored Everett
functions and the so-called Preisach representation theorem, see
\cite{Maye03} for detail. Different iterative inversion schemes,
which are equally suitable for a real-time feedforwarding, were
proposed, e.g. allowing also for an online adaptation in
\cite{tan2005}, and later in combination with an additional
observation of the recent hysteresis state in \cite{ruderman2010}.
The latter was also provided in a differential form, see the work
\cite{ruderman2016}, which is also used in the present
contribution.

Differing from the approaches mentioned above, we propose a new
type of the inversion-free feedforward hysteresis compensation.
Because the proposed hysteresis control scheme is generic, and
limited solely to the class of rate-independent hysteresis that
can be captured by the Preisach operator, no specific application
system is addressed explicitly. Notable possible applications,
however, are in the operation of electromagnets, piezoelectric and
magnetostrictive actuators, and other electro-magneto-mechanical
devices with a rate-independent hysteresis in the input channel.

The rest of the paper is structured as follows. The basic problem
formulation of a feedforward hysteresis compensation is given in
section \ref{sec:2}. Here we also provide an approximate
input-output view on the hysteresis behavior in terms of a
harmonic response. In section \ref{sec:3}, the Preisach hysteresis
operator is described, as far in detail as necessary, also in the
differential form used in the proposed control scheme. The phase
shift properties of the Preisach operator are elucidated based on
the extreme case of a lumped two-point switching hysteresis. The
proposed inversion-free hysteresis control scheme is introduced in
section \ref{sec:4}, together with analysis of the control errors
and achievable performance. The numerical examples of compensating
for one convex and one non-convex hysteresis are given in section
\ref{sec:5}. The paper is concluded by a discussion provided in
section \ref{sec:6}.

\section{Feedforward hysteresis control}
\label{sec:2}

\begin{figure}[!h]
\centering
\includegraphics[width=0.49\columnwidth]{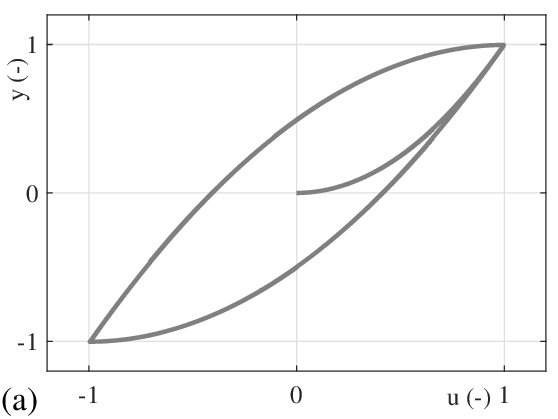}
\includegraphics[width=0.49\columnwidth]{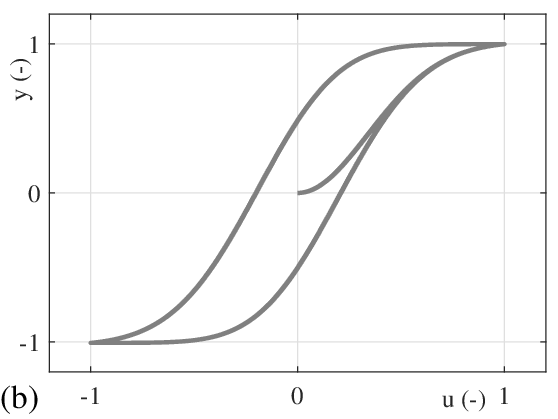}\\[0.5mm]
\includegraphics[width=0.49\columnwidth]{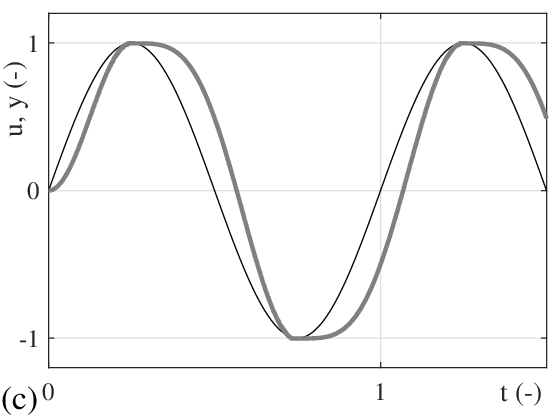}
\includegraphics[width=0.49\columnwidth]{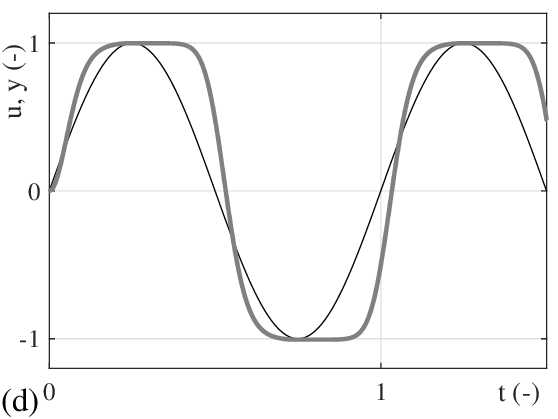}
\caption{Examples of a convex (a), (c) and non-convex (b), (d)
hysteresis; black thin line of input and grey thick line of output
in (c) and (d)} \label{fig:2}
\end{figure}

We consider a generic hysteresis system
\begin{equation}
y\bigl( t=\tau \bigr) = f\bigl[u(0 \leq t \leq \tau ),y_0 \bigr]
\label{eq:hyster}
\end{equation}
as a multi-valued nonlinear function $f[\cdot]$, which output at
the time $t=\tau$ depends on the recent and previous input values
at $t \leq \tau$ and the initial state $y_0 \equiv y(t=0)$.
Without loss of generality, we assume that $u(t) \in
\mathcal{C}^0$ and $y(t)$ is almost always piecewise
differentiable, except the reversal points where the
$\mathrm{sign}(\dot{u})$ changes. In other words, we do not allow
for stepwise inputs of hysteresis, while the input derivative does
not have to be continuous with respect to time. Recall that for
all input-output pairs $(u,y)$, cf. two examples in Fig.
\ref{fig:2} (a) and (b), one can say the system \eqref{eq:hyster}
possesses memory since at any instant $t$ the output $y(t)$ is
determined by the previous evolution of the input function $u(t)$,
and not only the input value at the same time instant, cf.
\cite{Visit94}.

\begin{figure}[!h]
\centering
\includegraphics[width=0.8\columnwidth]{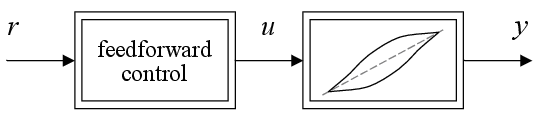}
\caption{Feedforward control of input-output hysteresis system}
\label{fig:1}
\end{figure}

Now the objective is to design a controller, see Fig. \ref{fig:1},
which will act as a feedforward hysteresis compensator and,
therefore, guarantee for $|y(t)-r(t)| < \varepsilon$, where $
\varepsilon > 0 $ is the smallest possible residual control error.
In addition, a potentially large control bandwidth is required,
that means $\varepsilon(j\omega)$ remains upper bounded for a
bounded angular frequency $\omega < \omega_b$. The design of a
feedforward controller assumes that an accurate model of the
hysteresis system is available, while we emphasize that the
proposed controller is inversion-free, i.e. it does not require
constructing or approximating the inverse of \eqref{eq:hyster}.
Recall that a closed analytic form of the inverse of a
multi-valued hysteresis function is not always available, like in
case of the Preisach hysteresis operator. At the same time,
feedforward hysteresis compensation may be required in several
applications where a hysteresis output $y$ appears rather as an
internal state and is not available (or can not be used) during
the controlled system operation.

As we focus on a rate-independent hysteresis in the strict sense
(for definition and properties of rate-independence of hysteresis
we reefer to e.g. \cite{Visit94}), the output $y$ depends on the
sequence of the input values but not on the frequency with which
they are proceeded. In other words, the rate-independent behavior
of a hysteresis system is not influenced by any affine
transformation on the time axis. It is also worth mentioning that
a suitable scaling, correspondingly normalization, of the
hysteresis input and output enables an assumption of the unity
domain and range of \eqref{eq:hyster}, cf. Fig. \ref{fig:2}. It
means that a simple static mapping $y(t)=u(t)$ would occur when
there is no hysteresis. Considering two illustrative examples,
depicted in Fig. \ref{fig:2} for a convex (a), (c) and non-convex
(b), (d) hysteresis, one can recognize that the output exhibits a
state-varying phase shift $\phi\bigl(u(t)\bigr)$ to a harmonic
input process. As originated from an Ancient Greek word
'hysteresis', meaning 'lagging behind', the most of the phase
shifts are of a lag type. That means, a time integral of the phase
shift over a full period has always a negative sign. At the same
time, a transient lead-type phase shift occurs for the non-convex
hysteresis, cf. Fig. \ref{fig:2} (d).

In order to analyze the input-output hysteresis properties, in a
more familiar sense of the dynamic system, one can approximate the
hysteresis output response to a harmonic input $u(t) = \sin(\omega
t)$ as follows
\begin{equation}
y(t) \approx  \bar{y}(u_0,y_0) + A\bigl(u(t)\bigr)  \sin
\Bigl(\omega t
 + \phi\bigl(u(t)\bigr) \Bigr). \label{eq:outharmonic}
\end{equation}
The bias $\bar{y}$ depends on the initial state of the hysteresis
system and reveals, this way, a signature of a hysteresis memory.
Generally, one can assume that $\bar{y}$ is bounded by the main
(major) loop of the hysteresis. Both, the gain factor $A(\cdot)$
and phase shift $\phi(\cdot)$ depend on the instantaneous
hysteresis state, while $|A| < \kappa < \infty$ and $-\pi/2 < \phi
< \pi/2$ can be assumed without loss of generality. The first
assumption is because the bounded gain follows immediately from
the Lipschitz-continuity of a hysteresis operator
\eqref{eq:hyster}, cf. \cite{Visit94}, while the corresponding
Lipschitz constant is $\kappa = \max \, \partial y /
\partial u$. The boundedness of the phase shift follows from the
input-output behavior of an elementary hysteresis operator
\emph{hysteron}, as demonstrated below in section \ref{sec:3}.

Now we are in the position to introduce the feedforward control
$u(t) = g\bigl[  r(t) \bigr]$, so that its serial connection with
a hysteresis system results in $y = f \bigl[ g[r], y_0 \bigr]
\rightarrow r$. Before doing it in section \ref{sec:4}, we will
first describe the Preisach hysteresis operator, which is assumed
for modeling the hysteresis system and used in the proposed
control scheme.

\section{Preisach hysteresis operator}
\label{sec:3}

The scalar Preisach hysteresis model \cite{Preis35,Maye03} is one
of the most powerful approaches for describing a multi-valued
rate-independent hysteresis mapping in its proper sense. The
Preisach hysteresis operator and its numerous extensions have been
used for several decades (see e.g. in \cite{BertMayer06}) in
magnetism, material science, but also in the control and system
engineering. We will briefly summarize the Preisach operator, as
far as necessary for our developments, while we refer to the
seminal literature \cite{Visit94,Maye03} for more fundamental and
profound details on mathematical hysteresis operators.

\begin{figure}[!h]
\centering
\includegraphics[width=0.5\columnwidth]{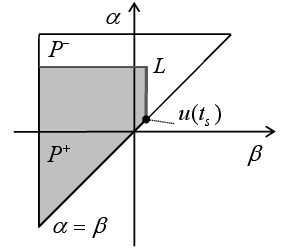}
\caption{Preisach plane with switching hysteresis state}
\label{fig:3}
\end{figure}

\begin{figure}[!h]
\centering
\includegraphics[width=0.46\columnwidth]{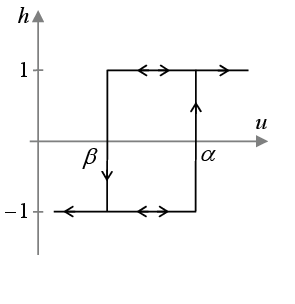}
\includegraphics[width=0.48\columnwidth]{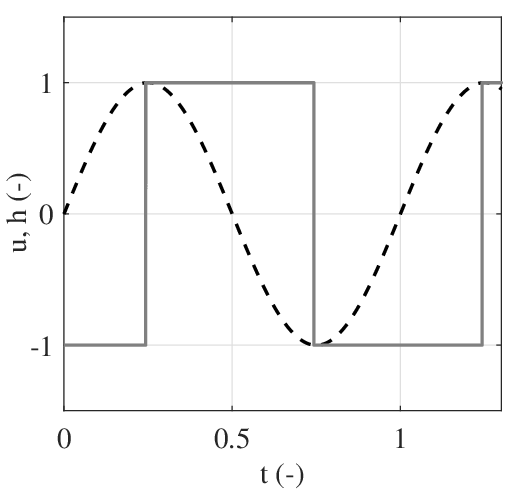}
\caption{Elementary hysteresis operator hysteron (on the left),
and the corresponding input- and output-value harmonics (on the
right)} \label{fig:hysteron}
\end{figure}

The memory affected multi-valued static map of the Preisach
hysteresis operator is given by
\begin{equation}
y(t) = \mathrm{H} \, [u(t)] = \iint\limits_{\alpha \geq \beta}
\rho(\alpha,\beta) \, h(\alpha,\beta)[u(t)] \mathrm{d}\alpha
\mathrm{d}\beta \,, \label{eq:preisach}
\end{equation}
in which the elementary nonlinear operator $h[\cdot]$ (also called
\emph{hysteron}) captures the spatially distributed internal state
of the corresponding input-output system. The hysteron is nothing
but an amplitude-delayed relay (see Fig. \ref{fig:hysteron} on the
left) which is parameterized by two threshold values $\alpha \geq
\beta$. Upon passing the threshold values, the output of $h[u]$
flips among the binary states $+1$ (up state) and $-1$ (down
state), for $u \geq \alpha$ and $u \leq \beta$ correspondingly.
For $\beta < u(t) < \alpha$, the hysteron keeps its previous
binary state for $ \forall \, t > t_s$ where $t_s$ is the time of
the last flipping, i.e. switching at the threshold value. The
entire Preisach operator is parameterized by the so-called
Preisach density function $\rho(\alpha,\beta)$, which is defined
over $P  = \{(\alpha,\beta) \; | \; \alpha \geq \beta\}$. The most
suitable geometrical interpretation of $P$ and, correspondingly,
state transitions of the Preisach hysteresis operator is by means
of the Preisach plane (further denoted by $P$). The plane is given
in the relative $(\alpha,\beta)$-coordinates of the input domain,
cf. Figs. \ref{fig:2} and \ref{fig:3}. It is worth saying in
advance that for a practical consideration and use of the Preisach
hysteresis operator, both the domain and range of
\eqref{eq:preisach} are bounded, so that
$$
u_{\min} \leq (\alpha,\beta) \leq u_{\max}
$$
and
$$
y_{\max} - y_{\min} = 2 \iint\limits_{P} \rho(\alpha,\beta) \,
\textrm{d}\alpha \textrm{d}\beta.
$$

At each time instant $t$, the Preisach plane is divided into two
disjunct subsets, i.e. $P(t)=P^{+}(t) \cup P^{-}(t)$, which
contain the hysterons in the 'up' state and 'down' state
correspondingly. Both subsets are separated by a staircase
interface $L$, cf. Fig. \ref{fig:3}, which represents the
instantaneous memory of the hysteresis system. The interface moves
from bottom to top for $\dot{u}>0$ and from right to left for
$\dot{u} <0$, while the dynamic transformations of $L$ occur by
including the new local extrema and deleting the previous,
according to the wiping-out hysteresis property, cf.
\cite{Maye03}.

One can easily show that
\begin{eqnarray}
\nonumber y(t) = & \iint\limits_{P^{+}} \rho(\alpha,\beta)
\textrm{d}\alpha \textrm{d}\beta - \iint\limits_{P^{-}}
\rho(\alpha,\beta)  \textrm{d}\alpha \textrm{d}\beta =
\\[1mm]
= & \iint \limits_{P} \rho(\alpha,\beta) \textrm{d}\alpha
\textrm{d}\beta - 2  \iint\limits_{P^{-}} \rho(\alpha,\beta)
\textrm{d}\alpha \textrm{d}\beta, \label{eq:preisdiff1}
\end{eqnarray}
when considering, for instance, a decreasing input. Therefore, for
two consecutive values $u(t_2) < u(t_1)$ of a monotonically
decreasing input $\forall \: t_2 > t_1$, the output difference can
be obtained, from \eqref{eq:preisdiff1}, as
\begin{eqnarray}
\nonumber \Delta y = y(t_{2}) - y(t_{1}) = - 2
\iint\limits_{P^{-}(t_{2})} \rho(\alpha,\beta) \textrm{d}\alpha
\textrm{d}\beta + \\
+ \, 2 \iint\limits_{P^{-}(t_{1})} \rho(\alpha,\beta)
\textrm{d}\alpha \textrm{d}\beta  = -2 \iint\limits_{\Omega}
\rho(\alpha,\beta) \, \textrm{d}\alpha \textrm{d}\beta \:.
\label{eq:preisdiff2}
\end{eqnarray}
Here $\Omega \equiv P^{-}(t_{1}) \setminus P^{-}(t_{2}) $ is the
difference set of the switching region in $P$, i.e. where the
hysterons flipped down during the time between $t_1$ and $t_2$.
Obviously, the switching region depends on $\Delta u =
u(t_2)-u(t_1)$, and when taking the limiting value $\lim \Delta u
\rightarrow 0 \equiv \mathrm{d}u$ one can obtain, without loss of
generality, the differential form of \eqref{eq:preisach} as
\begin{equation}
\mathrm{d} y  = 2 \, \mathrm{sign} (\dot{u})
\iint\limits_{\Omega(\mathrm{d} u)} \rho(\alpha,\beta)
\textrm{d}\alpha \textrm{d}\beta \: . \label{eq:preisdiff3}
\end{equation}
Note that at the switching time $t_s$, the set $\Omega(\mathrm{d}
u)$ coincides with the most bottom horizontal segment of $L$ for $
\mathrm{sign} (\dot{u}) = 1$ and with the most right vertical
segment of $L$ for $ \mathrm{sign} (\dot{u}) = -1$, both for a
monotonically changing $u(t)$. If the input direction changes,
i.e. the hysteresis operator experiences a local extremum, then
$\Omega(\mathrm{d} u) = L\bigl(u(t_s)\bigr)$, cf. Fig.
\ref{fig:3}. Integrating \eqref{eq:preisdiff3} with respect to
$y$, one obtains the Preisach hysteresis output, provided the
initial $y(0)$ is known. More details on this differential form of
the Preisach operator, also in the state-space formulation, can be
found in \cite{ruderman2016}.

Before introducing the hysteresis control, let us examine the
phase shift characteristics of a hysteresis operator, cf.
\eqref{eq:outharmonic} and Fig. \ref{fig:2}, based on the
input-output behavior of a single hysteron depicted in Fig.
\ref{fig:hysteron}. It is evident that the half of period of the
output level coincides with zero-crossing of the input harmonic.
Furthermore, the output switching, where $h$ crosses zero and has
its zero-average, occurs at time instances of the input peaks.
This provides exactly $\pm \pi/2$ phase shift between the input
and output harmonics of a hysteron, while the phase sign depends
on the initial state of the hysteron i.e. $h(t=0)$. We also recall
that the $h$ hysteresis loop (as depicted in Fig.
\ref{fig:hysteron} on the left) represents a boundary case of the
Preisach operator $\mathrm{H}$, i.e. with a maximally possible
loop area enclosed by the vertical hysteresis transitions at
$u_{\max}$ for 'up' and $u_{\min}$ for 'down'. Note that within
the $P$ plane, this corresponds to the case when all hysterons are
located in the left upper corner $(\alpha, \beta) = (u_{\max},
u_{\min})$. In general, closer the hysterons are located to the
$\alpha= \beta$ diagonal of the Preisach plane, lower phase shift
is between the input $u$ and corresponding output $h$. This allows
concluding $\max|\phi| = \pi/2$, cf. section \ref{sec:2} and Figs.
\ref{fig:2} and \ref{fig:hysteron} (on the right).

\section{Inversion-free hysteresis control}
\label{sec:4}

\begin{figure}[!h]
\centering
\includegraphics[width=0.45\columnwidth]{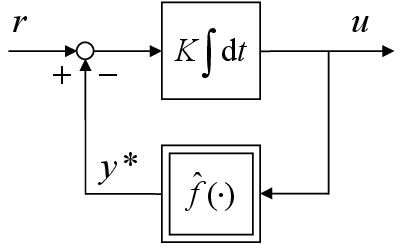}
\caption{Proposed feedforward hysteresis control $u=g[r]$}
\label{fig:control}
\end{figure}

The proposed inversion-free hysteresis control relies on an
internal model principle and incorporates the high-gain integral
feedback loop which aims controlling the hysteresis model
$\hat{f}[\cdot]$, see Fig. \ref{fig:control}.
If the internal model control loop achieves its goal, i.e. the
modeled hysteresis output $\hat{f}:\,  u(t) \mapsto y^*(t)$
follows the reference $r(t)$ as close as possible, then the output
of the integral regulator, which is
\begin{equation}
u(t) = K \int \bigl( r(t) - y^*(t) \bigr) \mathrm{d}t,
\label{eq:controllaw}
\end{equation}
mimics the inverse of the hysteresis map i.e. $\hat{f}^{-1} :\,
y^*(t) \approx r(t) \mapsto u(t)$. With this simple idea in mind,
let us analyze the bandwidth and accuracy of the hysteresis
controller $u=g[r]$, especially depending on the high-gain value
$K$. Since we are eager to see how close the internally controlled
$y^*(t)$ value follows the reference $r(t)$, we consider the
closed-loop error $e=r-y^*$ as the principal measure of accuracy
of the feedforward hysteresis compensator.

\begin{figure}[!h]
\centering
\includegraphics[width=0.98\columnwidth]{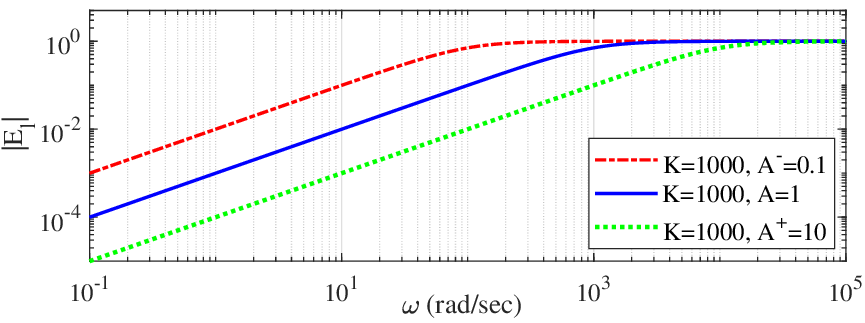}
\caption{Exemplary closed-loop error transfer function $E_{l}$ for
the control gain $K=1000$ and feedback variations: $A^{-}=0.1$,
$A=1$, $A^{+}=10$} \label{fig:frflin}
\end{figure}

Let us first examine the linear case, i.e. without hysteresis,
when a linear static mapping $y^* = A  u$ is in the closed-loop of
the $g$-compensator. The loop error transfer function is
\begin{equation}
E_l(j\omega) = \frac{e(j\omega)}{r(j\omega)} =
\frac{j\omega}{j\omega + K A}. \label{eq:error1}
\end{equation}
The above is a typical sensitivity function which drops towards
zero as $\omega \rightarrow 0$. Note that the corner frequency $(K
A)$ can be shifted to the right by increasing the control gain
$K$, provided the $A$-factor remains constant, cf. Fig.
\ref{fig:frflin}. When allowing for variations of $A(\cdot)$, that
is unavoidable for hysteresis nonlinearity, cf.
\eqref{eq:outharmonic}, the feedback loop remains stable provided
the static $A \neq \mathrm{const}$ nonlinearity satisfies the
sector condition, cf. e.g. \cite{khalil2002}. At the same time,
the sensitivity function becomes degraded by the low bound $A^{-}=
\min(A)$. This should be taken into account when designing $K$,
while the upper bound $A^{+}= \max(A)$ expands the control
bandwidth, cf. Fig. \ref{fig:frflin}.

Now, we expand our above consideration of the error transfer
function to the case of approximating hysteresis in the loop, cf.
\eqref{eq:outharmonic} and Fig. \ref{fig:control}. For the mostly
occurring lag-type phase shift $\phi$ (also most sensitive in
terms of the stability), we consider a standard first-order lag
transfer function
\begin{equation}
F(j\omega) = \frac{y^{*}(j\omega)}{u(j\omega)} = \frac{j\omega
/(\delta \omega_0) + 1}{j\omega \, \delta / \omega_0 +1}.
\label{eq:lag}
\end{equation}
Two parameters, $\omega_0$ and $\delta$, are characteristic for
approximating the hysteresis response to a harmonic input.
Obviously, due to the phase shift characteristics of interest, the
angular frequency $\omega_0$, at which the phase lag is maximal,
coincides with that of the harmonic input $u(j\omega_0)$.
\begin{figure}[!h]
\centering
\includegraphics[width=0.98\columnwidth]{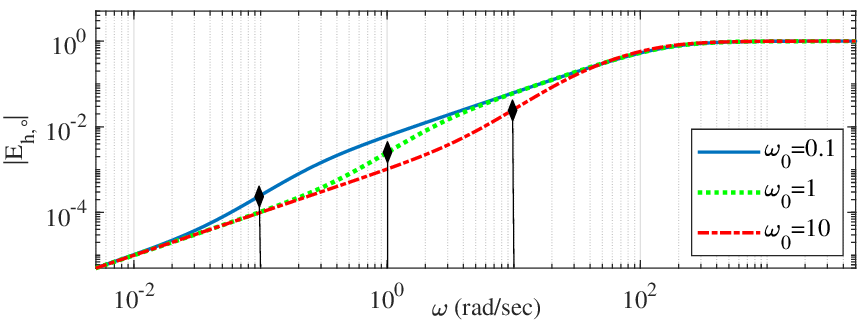}
\caption{Exemplary closed-loop error transfer function
$E_{h,\circ}$ for the assumed $K=1000$, $A=1$, and $\omega_0 =
\{0.1,\,1,\,10\}$ rad/sec} \label{fig:frfhyst}
\end{figure}
The factor $\delta$ scales the bandwidth of the lagging transfer
characteristics around $\omega_0$ and, thus, the phase-lag itself.
For the transfer function \eqref{eq:lag}, the phase lag is always
$-\pi/2 < \angle F(j\omega) < 0$. And for an average phase lag
$\angle F(j\omega) = - \pi/4$, which is sufficient for our
qualitative analysis, the $\delta = 2.5$ can be assumed in the
following. Substituting the approximation \eqref{eq:outharmonic}
instead of $\hat{f}(\cdot)$ into the closed control loop from Fig.
\ref{fig:control}, one obtains
\begin{equation}
e(j \omega) \Bigl(1 + \frac{K A}{j \omega} \, F (j \omega) \Bigr)
= r (j \omega) - \bar{y}. \label{eq:hysterror}
\end{equation}
It becomes evident that two transfer functions, $E_{h,r}(j \omega)
= e(j \omega)/r(j \omega)$ and $E_{h,\bar{y}}(j \omega) = e(j
\omega)/\bar{y}(j \omega)$, can be considered and analyzed
separately and in a similar manner. It is also worth recalling
that $\bar{y}$ constitutes the bias depending on the hysteresis
memory state, cf. \eqref{eq:outharmonic}. Therefore, only
steady-state or low-frequency range, i.e. $\omega \rightarrow 0$,
are relevant when analyzing $E_{h,\bar{y}}(j \omega)$. On the
contrary, one is focusing on the frequency range around $\omega =
\omega_0$, with a corresponding adjustment of the lag transfer
function \eqref{eq:lag}, when analyzing $E_{h,r}(j \omega)$. This
refers to the reference input $r(j \omega_0)$. Assuming, as
before, an exemplary control gain $K=1000$ and $A=1$, the
frequency response function of $E_{h,\circ}$ is shown in Fig.
\ref{fig:frfhyst} for $\omega_0 = \{0.1,\,1,\,10\}$ rad/sec. Note
that the analysis of $E_{h,\circ}$ transfer function is equally
valid in both cases $\circ = r \, \vee \, \bar{y}$. It is easy to
see, from the labeled black vertical bars, how the closed-loop
error changes depending on the angular frequency $\omega_0$ of the
reference signal. It can also be seen that for a steady-state
bias, injected through $\bar{y}$, the closed-loop error goes to
zero. Therefore, it becomes conclusive that with an increasing
$K$-value, assigned with respect to $A^{-}$ and $A^{+}$ (both are
known from the modeled hysteresis system), the same upper bound
for $|e|$ and, hence, for the hysteresis compensation error
$\varepsilon$, can be guaranteed despite an increasing $\omega_0$.
It implies, theoretically, an infinite control bandwidth, provided
an infinite $K$-gain and an appropriate modeling of the hysteresis
function $\hat{f}$ are realizable.

\section{Numerical evaluation}
\label{sec:5}

The proposed control is evaluated numerically together with the
hysteresis system plant, both interconnected as depicted in Fig.
\ref{fig:1}. The simulated hysteresis system $f[\cdot]$ and the
one involved in the designed control, i.e. $\hat{f}[\cdot]$, use
the same Preisach hysteresis operator in the differential form,
see section \ref{sec:3}. The discrete $(\beta,\alpha)$-mesh of the
$400 \times 400$ size is assumed, that results in a total of 80200
hysterons, cf. \cite{ruderman2018}. Note that the allocated
$(\beta,\alpha)$-matrix represents simultaneously the parameters
space of the Preisach density function and the state-space of the
binary hysterons. This allows for a memory- and
computation-efficient implementation of the Preisach model, which
is equally suitable for (and has been tested in) a real-time
environment. Further details on the discretized real-time
implementation in the differential form can be found in
\cite{ruderman2010}. The high-gain of the internal model loop is
assigned as $K=6000$. This is done with respect to the input
frequency, on the one hand, and the discretization level of the
Preisach operator, on the other hand. Recall that the latter
provokes a finite quantization of $y$ and $y^*$ and, thus,
violates the theoretical assumption of the hysteresis output to be
piecewise differentiable between two reversal points.

\begin{figure}[!h]
\centering
\includegraphics[width=0.49\columnwidth]{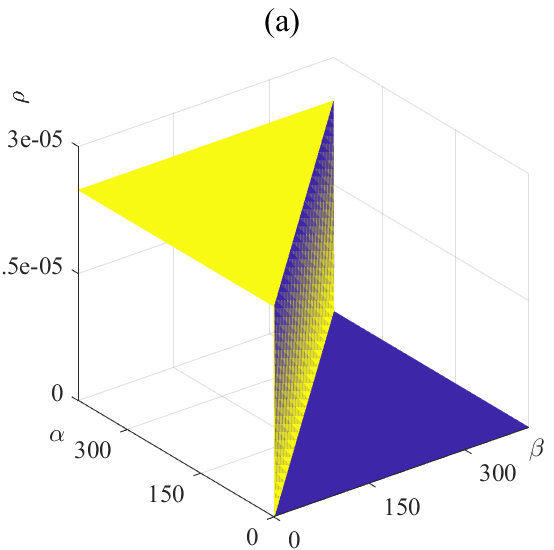}
\includegraphics[width=0.49\columnwidth]{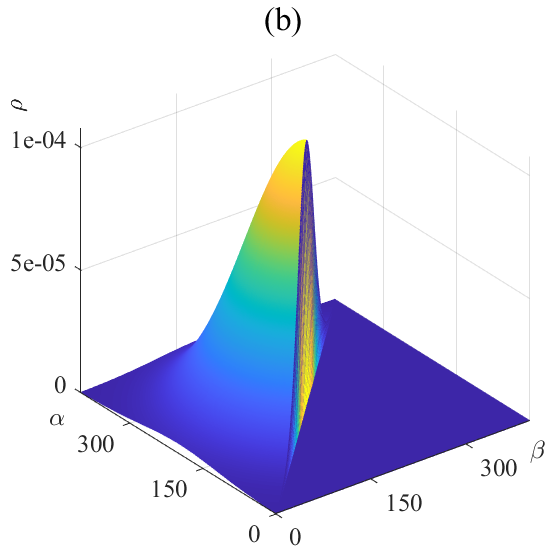}
\caption{Assumed Preisach density function over the discretized
$(\beta,\alpha)$ plane, (a) uniform distribution, (b) Gaussian
normal distribution} \label{fig:plane}
\end{figure}

Two Preisach density functions depicted in Fig. \ref{fig:plane}
are exemplary taken for evaluation. The first one, shown in (a),
is a uniform distribution of the hysterons with
$\rho(\alpha,\beta)=\mathrm{const}$. The second one, shown in (b),
is a two-dimensional Gaussian normal distribution
$\mathcal{N}(\mu, \Sigma)$, which is parameterized by the mean
vector $\mu$ and symmetric covariance matrix $\Sigma$, both in the
relative $(\alpha,\beta)$ coordinates. In both $\rho$-cases, the
relative weights of the hysterons have been assumed so that the
saturated hysteresis has the range $y_{\min,\max} = \{-1, 1\}$,
while the input domain is considered to be of the same scale $-1
\leq u \leq 1$. Note that the uniform and Gaussian normal Preisach
density functions result in the hysteresis loops depicted in Fig.
\ref{fig:2} (a) and (b), correspondingly.

First, a 'zigzag'-shaped input reference with a continuously
increasing amplitude was applied for both Preisach density
functions. This results in a set of continuously growing
hysteresis loops, which are enveloped inside of each other.
\begin{figure}[!h]
\centering
\includegraphics[width=0.48\columnwidth]{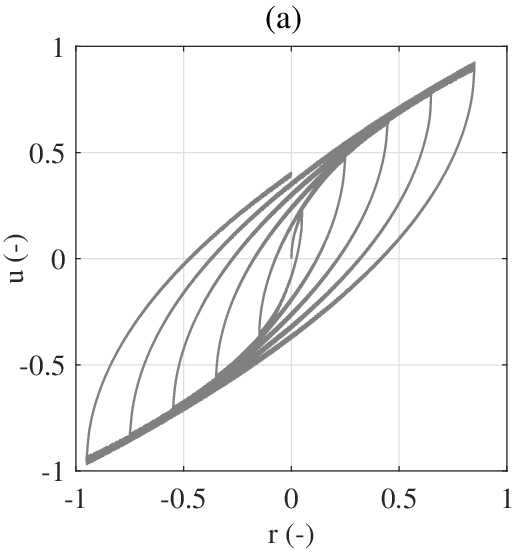}
\includegraphics[width=0.48\columnwidth]{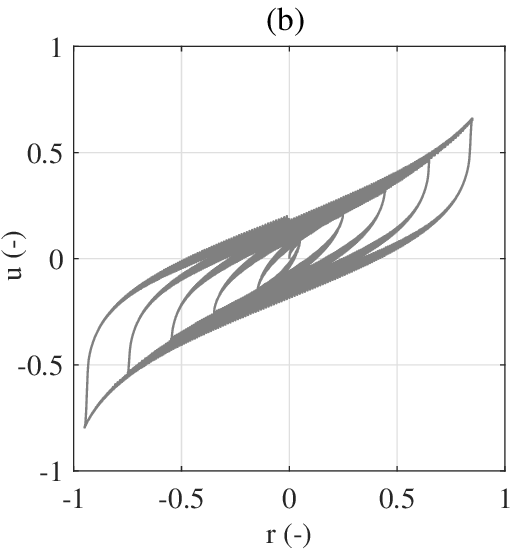}\\[1mm]
\includegraphics[width=0.48\columnwidth]{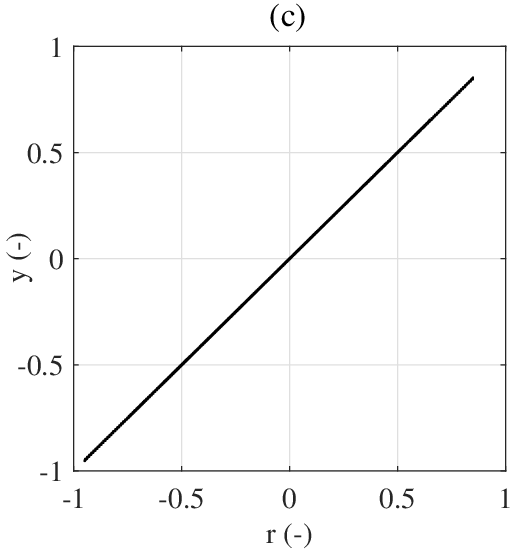}
\includegraphics[width=0.48\columnwidth]{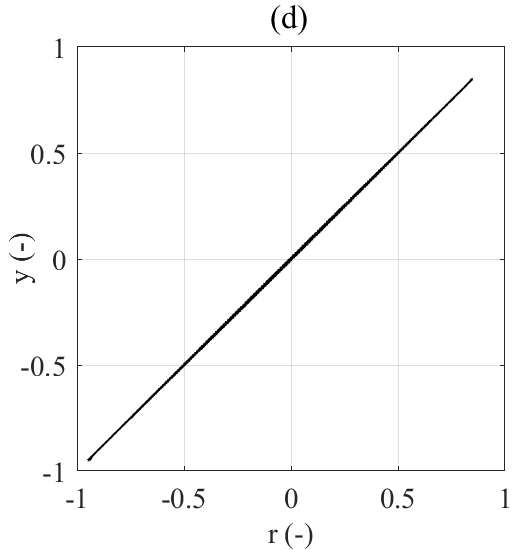}
\caption{Control over reference values (above) and output over
reference values (below) of the hysteresis compensation for a
zigzag shaped input, (a) and (c) uniform distribution, (b) and (d)
Gaussian normal distribution} \label{fig:comp1}
\end{figure}
The produced control signal is shown in Fig. \ref{fig:comp1} in
the $(r,u)$ coordinates, for the uniform in (a) and for the
Gaussian normal Preisach density function in (b), respectively.
The compensated input-output behavior is shown in Fig.
\ref{fig:comp1} in the $(r,y)$ coordinates, in (c) and (d)
correspondingly.

\begin{figure}[!h]
\centering
\includegraphics[width=0.48\columnwidth]{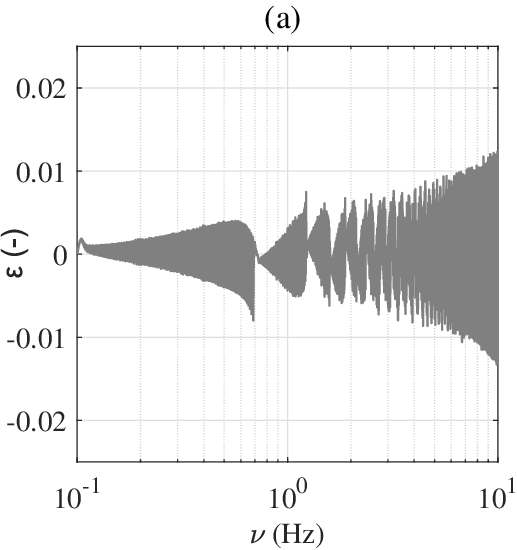}
\includegraphics[width=0.48\columnwidth]{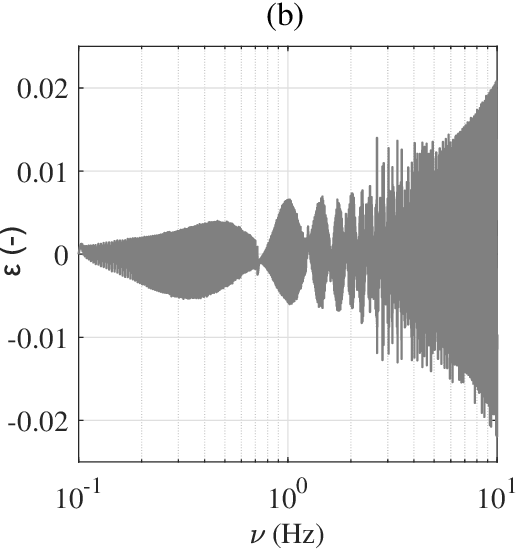}\\[1mm]
\includegraphics[width=0.48\columnwidth]{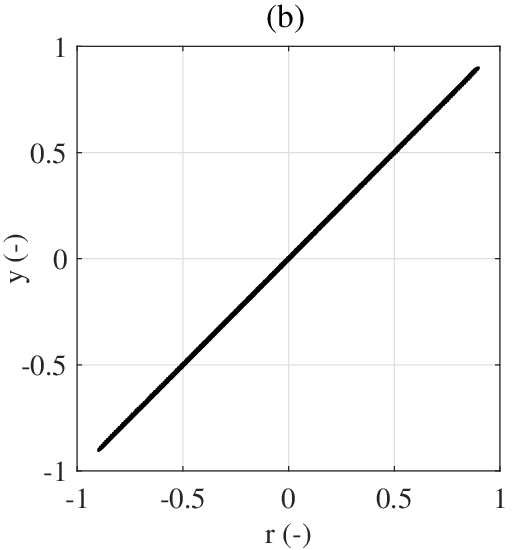}
\includegraphics[width=0.48\columnwidth]{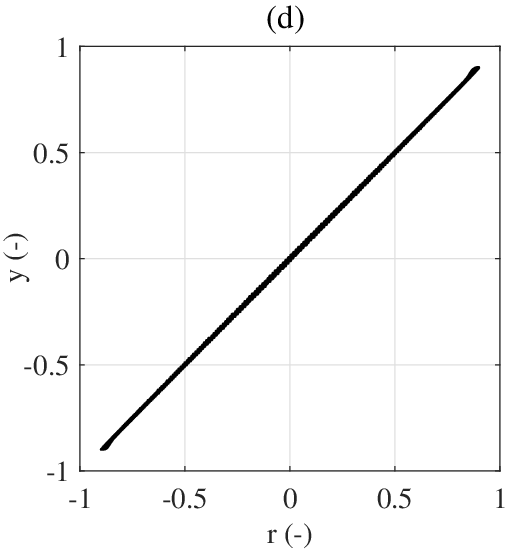}
\caption{Control error $\varepsilon(\nu)$ (above) and output over
reference values (below) of the hysteresis compensation for a
0.1-10 Hz chirp input, (a) and (c) uniform distribution, (b) and
(d) Gaussian normal distribution} \label{fig:comp2}
\end{figure}

Next, a chirp reference input $u = 0.9 \sin \bigl( (2\pi \nu t)t
\bigr)$, was applied with a linearly progressing frequency $\nu t$
between 0.1 and 10 Hz. Note that the reference amplitude was set
to 0.9 for not reaching a fully saturated hysteresis state. The
compensation results are shown in Fig. \ref{fig:comp2}, for the
uniform Preisach density function on the left and for the Gaussian
normal Preisach density function on the right. The hysteresis
compensation error $\varepsilon(\nu)$ is depicted as a function of
the growing reference frequency in (a) and (b). The compensated
input-output behavior is depicted in Fig. \ref{fig:comp2} in the
$(r,y)$ coordinates, in (c) and (d) correspondingly.

\section{Discussion}
\label{sec:6}

This paper proposed an inversion-free feedforward control of
hysteresis systems, based on the internal model principle and
Preisach operator of an arbitrary rate-independent hysteresis map.
The Preisach model, which relies on a phenomenon-based weighted
superposition of multiple elementary hysteresis operators
\emph{hysterons}, is the most flexible among the operator-based
hysteresis models for shaping the hysteresis loops and the
associated hysteresis memory. Although the existence of an inverse
Preisach operator has been proven mathematically
\cite{BrokateVisintin89}, there is no sound piecewise continuous
and implementable analytical form for it up to now. Therefore, the
proposed alternative inversion-free feedforward hysteresis
compensator, which is based on the same standard Preisach
operator, is advantageous. The single design parameter, apart from
the identified Preisach model of the hysteresis system to be
controlled, is the high-gain of an integral loop. For
identification and adaptation of the Preisach hysteresis model
itself, we refer to the lately proposed robust online estimator of
the Preisach density function, see \cite{ruderman2018}. Since the
high integral gain is used for an internal model control loop only
and, thus, does not produce any saturated control actions, its
value can be increased (at least theoretically) towards infinity.
This results in a theoretically \emph{infinite bandwidth} of the
compensator. This was analyzed in section \ref{sec:4} and
evaluated with numerical examples in section \ref{sec:5}. It was
also shown that the upper bound of the residual control error is
growing with 20 dB/dec  in frequency domain of the reference
input, cf. Figs. \ref{fig:frfhyst} and \ref{fig:comp2} (a) and
(b). However, this fact represents a rather implementation- and
application-related issue. Apart from a theoretically justified
hysteresis compensation error, the residual error contents in
$\varepsilon(t)$ (propagated from $e(t)$), are of a numerical
nature owing to the discretized Preisach hysteresis operator.
Justified, a finite $(\alpha,\beta)$ mesh produces a quantized, to
say 'staircase'-type, output $y^*$ of the Preisach model. As this
signal is fed back in a high-gain integral loop, the associated
parasitic side-effects become visible in the resulted $(r,u)$
mapping, cf. Fig. \ref{fig:comp1} (a), (b). A possible solution is
increasing the size of $(\alpha,\beta)$-mesh with respect to
$\omega_0$ and $K$-gain values or using an $y^*$-interpolator.

\bibliographystyle{IEEEtran}
\bibliography{references}

\begin{thebibliography}{10}
\providecommand{\url}[1]{#1}
\csname url@rmstyle\endcsname
\providecommand{\newblock}{\relax}
\providecommand{\bibinfo}[2]{#2}
\providecommand\BIBentrySTDinterwordspacing{\spaceskip=0pt\relax}
\providecommand\BIBentryALTinterwordstretchfactor{4}
\providecommand\BIBentryALTinterwordspacing{\spaceskip=\fontdimen2\font plus
\BIBentryALTinterwordstretchfactor\fontdimen3\font minus
  \fontdimen4\font\relax}
\providecommand\BIBforeignlanguage[2]{{%
\expandafter\ifx\csname l@#1\endcsname\relax
\typeout{** WARNING: IEEEtran.bst: No hyphenation pattern has been}%
\typeout{** loaded for the language `#1'. Using the pattern for}%
\typeout{** the default language instead.}%
\else
\language=\csname l@#1\endcsname
\fi
#2}}

\bibitem{BertMayer06}
G.~Bertotti and I.~D. Mayergoyz, \emph{The Science of Hysteresis 1--3},
  1st~ed.\hskip 1em plus 0.5em minus 0.4em\relax Academic Press, 2006.

\bibitem{tao1995}
G.~Tao and P.~V. Kokotovic, ``Adaptive control of plants with unknown
  hystereses,'' \emph{IEEE Transactions on Automatic Control}, vol.~40, no.~2,
  pp. 200--212, 1995.

\bibitem{esbrook2012}
A.~Esbrook, X.~Tan, and H.~K. Khalil, ``Control of systems with hysteresis via
  servocompensation and its application to nanopositioning,'' \emph{IEEE
  Transactions on Control Systems Technology}, vol.~21, no.~3, pp. 725--738,
  2012.

\bibitem{gorbet2001}
R.~B. Gorbet, K.~A. Morris, and D.~W. Wang, ``Passivity-based stability and
  control of hysteresis in smart actuators,'' \emph{IEEE Transactions on
  Control Systems Technology}, vol.~9, no.~1, pp. 5--16, 2001.

\bibitem{cruz2001}
J.~M. Cruz-Hern{\'a}ndez and V.~Hayward, ``Phase control approach to hysteresis
  reduction,'' \emph{IEEE transactions on Control Systems Technology}, vol.~9,
  no.~1, pp. 17--26, 2001.

\bibitem{vasquez2020}
M.~Vasquez-Beltran, B.~Jayawardhana, and R.~Peletier, ``Recursive algorithm for
  the control of output remnant of {Preisach} hysteresis operator,'' \emph{IEEE
  Cont. Sys. Letters}, vol.~5, no.~3, pp. 1061--1066, 2020.

\bibitem{Janaideh2018}
M.~Al~Janaideh, M.~Rakotondrabe, I.~Al-Darabsah, and O.~Aljanaideh, ``Internal
  model-based feedback control design for inversion-free feedforward
  rate-dependent hysteresis compensation of piezoelectric cantilever
  actuator,'' \emph{Control Engin. Practice}, vol.~72, pp. 29--41, 2018.

\bibitem{Visit94}
A.~Visintin, \emph{Differential Models of Hysteresis}, 1st~ed.\hskip 1em plus
  0.5em minus 0.4em\relax Springer, 1994.

\bibitem{krejci2001}
P.~Krejci and K.~Kuhnen, ``Inverse control of systems with hysteresis and
  creep,'' \emph{IEE Proceedings - Control Theory and Applications}, vol. 148,
  no.~3, pp. 185--192, 2001.

\bibitem{Preis35}
F.~Preisach, ``Ueber die magnetische {Nachwirkung},'' \emph{Zeitschrift
  Physik}, vol.~94, pp. 277--302, 1935.

\bibitem{BrokateVisintin89}
M.~Brokate and A.~Visintin, ``Properties of the {Preisach} model for
  hysteresis,'' \emph{Journal f{\"u}r die reine und angewandte Mathematik
  (Crelles Journal)}, no. 402, pp. 1--40, 1989.

\bibitem{davino2005}
D.~Davino, C.~Natale, S.~Pirozzi, and C.~Visone, ``A fast compensation
  algorithm for real-time control of magnetostrictive actuators,''
  \emph{Journal of Magnetism and Magnetic Materials}, vol. 290, pp. 1351--1354,
  2005.

\bibitem{Maye03}
I.~D. Mayergoyz, \emph{Mathematical models of hysteresis and their
  application}, 2nd~ed.\hskip 1em plus 0.5em minus 0.4em\relax Academic Press
  (imprint of Elsevier), 2003.

\bibitem{tan2005}
X.~Tan and J.~S. Baras, ``Adaptive identification and control of hysteresis in
  smart materials,'' \emph{IEEE Transactions on Automatic Control}, vol.~50,
  no.~6, pp. 827--839, 2005.

\bibitem{ruderman2010}
M.~Ruderman and T.~Bertram, ``Discrete dynamic {Preisach} model for robust
  inverse control of hysteresis systems,'' in \emph{IEEE 49th Conference on
  Decision and Control (CDC)}, 2010, pp. 3463--3468.

\bibitem{ruderman2016}
M.~Ruderman, ``State-space formulation of scalar {Preisach} hysteresis model
  for rapid computation in time domain,'' \emph{Applied Mathematical
  Modelling}, vol.~40, no.~4, pp. 3451--3458, 2016.

\bibitem{khalil2002}
H.~Khalil, \emph{Nonlinear Systems}, 3rd~ed.\hskip 1em plus 0.5em minus
  0.4em\relax Prentice Hall, 2002.

\bibitem{ruderman2018}
M.~Ruderman and D.~Rachinskii, ``Discrete-time adaptive hysteresis filter for
  parallel computing and recursive identification of {Preisach} model,'' in
  \emph{IEEE Conference on Control Technology and Applications (CCTA)}, 2018,
  pp. 1096--1101.

\end{thebibliography}

\end{document}